\documentclass[a4paper]{amsart}
\usepackage[usenames]{color}
\usepackage{epsfig}

\newtheorem{theorem}{Theorem}[section]
\newtheorem{corollary}[theorem]{Corollary}
\newtheorem{lemma}[theorem]{Lemma}

\newtheorem{remarks}[theorem]{Remarks}

\title{$H_2 $ Molecule in Strong Magnetic Fields }

\author{M. Beau$^1 $, R. Benguria$^2 $,
R. Brummelhuis$^3 $, P. Duclos$^4 $}

\thanks{The work of Rafael Benguria has been supported by ICM (Chile), grant
P07--027--F.}
\address{$^{1,4} $ Centre de Physique Th\'eorique UMR 6207 - Unit\'e Mixte de
Recherche du CNRS et des Universit\'es Aix-Marseille I,
Aix-Marseille II et de l' Universit\'e du Sud Toulon-Var -
Laboratoire affili\'e \`a la FRUMAM, Luminy Case 907, F-13288
Marseille Cedex 9 France.}\address{$^2 $ Departamento de F\'\i
sica, P. Universidad Cat\'olica de Chile, Santiago, Chile.}

\address{$^3 $ Department of Economics, Mathematics and
Statistics, Birkbeck, University of London, Malet Street, WC1E 7HX
London, United Kingdom and Laboratoire de Math\'ematiques de
Reims, EA 4535.}

\begin{document}

\maketitle

\begin{abstract}
The Pauli-Hamiltonian of a molecule with fixed nuclei in a strong
constant magnetic field is asymptotic, in norm-resolvent sense, to
an effective Hamiltonian which has the form of a multi-particle
Schr\"odinger operator with interactions given by one-dimensional
$\delta $-potentials. We study this effective Hamiltonian in the
case of the $H_2 $-molecule and establish existence of the ground
state. We also show that the inter-nuclear equilibrium distance
tends to 0 as the field-strength tends to infinity.
\end{abstract}

\section{\bf Introduction}

In the final decade before his untimely death, Pierre Duclos
embarked upon a large--scale systematic study of atoms and
molecules in strong homogeneous magnetic fields. This topic had
already received considerable attention in the astrophysics and
mathematical physics community (see for example \cite{AtMol},
\cite{La}, \cite{NeuSt} and \cite{LSY}). The published record of
Pierre's involvement started with the paper \cite{BD1} (which can
to a large measure be accredited to him), which established the
norm-resolvent convergence, as the magnetic field strength tends
to infinity, of the Pauli Hamiltonian of a non-relativistic atom
in a constant magnetic field to an effective Hamiltonian
describing a one-dimensional atom with the electrostatic
interaction replaced by Dirac $\delta $ point-interactions. Though
there exists a long and respected tradition in theoretical physics
and chemistry of using Hamiltonians with $\delta $-potentials as
toy models (going back, at least, to the Kronig and Penney model
of solid state physics), this theorem gave an example of such
$\delta $-potentials naturally arising as part of a genuine
physical problem. Around the same time, \cite{BaSoY} proved by a
different method that the ground-state energy of the true
Hamiltonian converges to that of the corresponding $\delta
$-Hamiltonian. The Hartree functional associated to the $\delta
$-Hamiltonian had already made its appearance in \cite{LSY}.

The approach initiated in \cite{BD1} was further developed in
\cite{BD3}, which established a three-member hierarchy of
effective multi-particle Hamiltonians on the line, each one of
which is asymptotic, in norm-resolvent sense, to the full
Pauli-Hamiltonian, but with increasingly better rates of
convergence. The $\delta $-Hamiltonian is the member of this
family with the slowest convergence. This paper also analyzed the
effect of imposing particle anti-symmetry which, on the level of
the $\delta $-Hamiltonian, translates into the latter acting on a
certain direct sum of fermionic $L^2 $-spaces and $L^2 $-spaces
with no symmetry-restriction at all; cf. section 2 below for the
case of two electrons. An early announcement of some of the
results of \cite{BD3}, which took a long time in writing, was made
in \cite{BD2}.

Despite its apparent simplicity, the spectral analysis of these
$\delta $-Hamiltonians
is far from trivial, except in the case of a single electron. As a
first step, Pierre, together with Santiago P\'erez-Oyarz\'un and
two of the present authors, investigated in \cite{BBDP} the
molecular $H_2 ^+ $-ion in the Born-Oppenheimer approximation in
strong magnetic fields, taking the corresponding $\delta $-model
as its starting point. The latter is explicitly solvable for a one
electron molecule, and its solution can be used, in combination
with second order perturbation theory, to obtain the correct
values of the equilibrium distance and of the binding energy for
strong magnetic fields of the order of $10^9 $ to $10^{14 } . $
a.u. Building upon this paper, it was shown in \cite{BBDPV1} (see
also \cite{BBDPV2}) that as the field-strength $B $ increases, the
equilibrium distance between the nuclei in the Born-Oppenheimer
approximation tends to 0, both for the $\delta $-model and for the
full molecular Pauli Hamiltonian, for the latter at a rate of
$(\log B)^{-3/2 } . $ This naturally suggested the possibility of
using strong magnetic fields to facilitate nuclear fusion by
enhanced tunnelling through the Coulomb barrier, a point which was
taken up by Ackermann and Hogreve \cite{AH}, who numerically
computed accurate equilibrium separation distances, as well as the
corresponding nuclear fusion rates, for a range of $B $'s from 0
and $10^4 . $ For field strengths $B $ of the order of $10^4 $
they indeed found a drastic increase of the tunnelling cross
sections. Unfortunately, such $B $'s are still much too big to be
generated in a laboratory on earth.

In a parallel set of papers \cite{CDR} and \cite{BBDR} (see also
\cite{B}), as well as in hitherto unpublished work, Pierre, in
cooperation with various co-authors and building upon an important
earlier paper of Rosenthal \cite{Ros}, developed a systematic
method for analyzing the spectrum of these multi-particle $\delta
$-Hamiltonians, which he baptised the skeleton method, and which
basically reduces the spectral analysis of such an Hamiltonian to
that of a finite-dimensional system of $(N - 1 ) $-dimensional
integral operators, $N $ being the number of particles (note that
$N - 1 $ is then the dimension of the support of any of the
$\delta $-potentials occurring in this $\delta $-Hamiltonian).
This system, which is called the skeleton of the original
Hamiltonian, is still too complicated to be solved explicitly even
for an $N $ as small as 2, but opens the way for a systematic
numerical approach (as was done by Rosenthal for the Helium-type
$\delta $-atom with $N = 2 .$). For arbitrary $N $ the size of
this system of integral operators will be $N (N + 1 )/2 $, though
this can be reduced by particle symmetry and parity
considerations.
\medskip

The present paper takes up the work of \cite{BBDPV1},
\cite{BBDPV2}  for the $H_2 ^+ $-ion by studying the existence and
equilibrium distance of the $H_2 $-molecule in the $\delta
$-approximation. We show that the equilibrium distance, in this
approximation, again tends to 0 as $B \to \infty $, at the same
speed as that found for $H_2 ^+ $. The paper should be seen as a
part of a larger and more ambitious project of Pierre, which was
to prove this for arbitrary Pauli molecules in strong magnetic
fields. In the context of enhanced fusion, one might hope,
following \cite{AH}, that adding more electrons would lead to
increased shrinking of the equilibrium distance through better
shielding of the nuclear charges. Our results suggest that such an
effect will not show up in the top order asymptotics of this
equilibrium distance as function of the field strength $B $,
though it might of course still make itself noticeable at finite
but large $B $. A major problem for the analysis of the $H_2
$-molecule is that, unlike for $H_2 ^+ $, the corresponding
$\delta $-model is no longer explicitly solvable, and we will
arrive at our conclusions by a combination of analytical and
numerical methods. In particular, a key intermediary result we
need is that the electronic ground-state energy (that is, leaving
out the internuclear repulsion term) of the molecular $\delta
$-Hamiltonian be an increasing function of the internuclear
distance. This seems physically intuitive, but we have been unable
to find an analytic proof. For non-relativistic single-electron
molecules without magnetic field, monotonicity of the electronic
energy in the internuclear distance was shown in \cite{LS},
\cite{L}; we are not aware of any rigorous results for the
multi-electron case, with or without magnetic field. By deriving
explicit estimates for the equilibrium distance and minimal energy
associated to an elementary variational upper bound of the
molecular ground-state energy, and combining these with a trivial
lower bound, we are able to show that it suffices to know the
monotonicity of the (true) electronic ground-state for a certain
range of internuclear distances. The latter was, amongst other
things, verified numerically in \cite{B}, using Pierre's skeleton
method.
\medskip

As will be clear, this paper's conclusions are far from complete,
and the paper is perhaps best seen as a work-in-progress report on
a project Pierre was working on at the time of his death, one of
the very many he was actively involved in. His absence, and the
impossibility of the lively exchange of ideas we had grown
accustomed to having with him, are sorely felt by the three
surviving authors.

\section{\bf Notation and Main Results}

As mentioned in the introduction, we want to study the equilibrium
distance, in the Born-Oppenheimer approximation, of the
non-relativistic $H_2 $-molecule in a strong constant magnetic
field, in the large field limit. We take the magnetic field
directed along the $z $-axis, and the two nuclei, of charge $Z $,
aligned in the direction of the field\footnote{The methods of
\cite{BD1}, \cite{BD3} on which this paper is based unfortunately
do not apply if the molecule is not aligned with the field, since
the angular momentum in the field direction is not conserved
anymore.} and located at $\pm \frac{1 }{2 } R \widehat{e }_z $,
where $\widehat{e }_z = (0, 0, 1 ) $ is the unit vector in the
direction of the $z $-axis. The molecule is described by the
familiar two-electron Pauli-Hamiltonian $\mathbb{H }^B $ (which we
won't write down explicitly) acting on anti-symmetric
wave-functions which include spin-coordinates, and therefore are
functions of $(r_i, s_i ) $, $i = 1, 2 $, where $r_i $ are the
spatial variables of the $i $-th electron, and $s_i = \pm 1 $ are
its the spin variables; the anti-symmetrisation is done with
respect to all the variables. We fix the total (orbital) angular
momentum in the field-direction to be $\mathbb{M } $, which we can
do since the corresponding component of the angular momentum
operator commutes with the Pauli-Hamiltonian. In this situation it
can be shown by the methods of \cite{BD1}, \cite{BD3} that the
full Hamiltonian $\mathbb{H }^B $, after projection onto the
lowest Landau band, is asymptotic, in norm-resolvent sense, to the
following two-electron Schr\"odinger operator with $\delta
$-potentials on $\mathbb{R }^2 $:
\begin{equation} \label{H-delta}
H_{\delta } := H_{\delta } (a, Z , \varepsilon ) := \sum _{i = 1 }
^2 \left( - \frac{1 }{2 } \Delta _{z_i } - \sum _{\pm } \delta
(z_i \pm a ) \right) + \frac{1 }{Z } \delta (z_1 - z_2 ) +
\frac{\varepsilon }{2a } ;
\end{equation}
$H_{\delta } $ will act on a certain direct sum of copies of $L^2
(\mathbb{R }^2 ) $ and $L^2 _{\rm a.s.} (\mathbb{R }^2 ) $ which
we specify below, where "a.s." stands for asymmetric
wave-functions. The parameters $a $ and $\varepsilon $ in the
definition of $H_{\delta } $ are related to the original
parameters $R, Z $ and $B $ by
\begin{equation} \label{def:a;epsilon}
a := R L Z / 2 , \ \ \varepsilon := Z / L ,
\end{equation}
where $L = L(B) = 2 W (\sqrt{B }/2 ) $, $W $ being the principal
branch of the Lambert-function, defined as that branch of the
inverse of $x e^x $ which passes through 0 and is positive for
positive $x $; cf., \cite{Lambert}. Note that $\varepsilon =
\varepsilon (B ) \to 0 $ as $B \to \infty $, since $L(B) \simeq
\log B $. We also let
\begin{equation} \label{h-delta}
h_{\delta } := h_{\delta } (a, Z ) := H_{\delta }  -
\frac{\varepsilon }{2a } ,
\end{equation}
the electronic part of $H_{\delta } $ (at fixed $a $). The
operators $H_{\delta } $ and $h_{\delta } $ are defined in
form-sense, as closed quadratic forms on the first order Sobolev
space $H^1 (\mathbb{R } ^2 ) $. Their operator domain consists of
those functions $\psi $ in $H^1 (\mathbb{R } ^2 ) $ whose
restrictions to $\mathbb{R }^2 \setminus \{ z_1 = z_2 , \, z_i =
\pm a \ \mbox{for} \ i = 1, 2 \} $ are in the second order Sobolev
space $H^2 $, and whose gradients satisfy an appropriate
jump--condition across the supports of the different $\delta
$-potentials: specifically $\partial _{z_i } \psi |_{z_i = \pm a -
} - \partial _{z_i } \psi |_{z_i = \pm a + } =  2 \psi |_{z_i =
\pm a } $, with a similar condition involving the normal
derivative of $\psi $ across $\{z_1 = z_2 \} $, but with opposite
sign: cf., the Appendix of \cite{BD3} for details.

We refer to \cite{BD3} for the precise technical sense in which
$\mathbb{H }^B $ and $H_{\delta } $ are asymptotic, but note that this will
imply that if $H_{\delta } $ has an eigenvalue $E(L) < 0
$ then $\mathbb{H }^B $ will have an eigenvalue at distance $O(L) $ of $L^2 Z^2 E(L) . $
\medskip

The Hilbert space in which $H_{\delta } $ and $h_{\delta } $ act
is , by \cite{BD3}, theorem 1.8,
\begin{equation}
L^2 (\mathbb{R }^2  ) ^{\# \mathcal{M }_1 } \oplus L^2 _{\rm a.s.
} (\mathbb{R }^2 )^{\# \mathcal{M }_2 } ,
\end{equation}
where $\mathcal{M }_1 = \{ (m_1 , m_2 ) \in \mathbb{N } \times
\mathbb{N } : m_1 + m_2 = \mathbb{M }, \ m_1 < m_2 \} $ and
$\mathcal{M }_2 = \{ (m , m ) : m \in \mathbb{N } , 2m =
\mathbb{M} \} $, which is either empty or a singleton (depending
on whether $\mathbb{M } $ is odd or even). Observe that unless
$\mathbb{M } \neq 0 $, $\mathcal{M }_1 $ will always be non-empty,
and, consequently, the infimum of the spectrum of $H_{\delta } $
will be the infimum of its spectrum on $L^2 (\mathbb{R }^2 ) $,
which equals the infimum of the spectrum of its restriction to the
bosonic subspace of symmetric wave-functions in $L^2 (\mathbb{R
}^2 ) $, despite having originally started off with fermionic
electrons. For the higher lying eigenvalues, no symmetry
restrictions have to be taken into account, at least for the
$\delta $-Hamiltonian - the situation will be different for the
full Pauli-Hamiltonian.

Since for $\mathbb{M } = 0 $, $H_{\delta } $ acts on $L^2 _{\rm
a.s.} (\mathbb{R }^2 ) $ while for $\mathbb{M } \geq 1 $, it acts
on a Hilbert-space containing $L^2 (\mathbb{R }^2 ) $ as a direct
summand, it is clear, for any fixed value of the inter-nuclear
distance $a $, that the ground state of $h_{\delta } $, if it
exists, occurs in any of the sectors with $\mathbb{M } \geq 1 $.
It then follows that for sufficiently large $B $, the ground state
of (the electronic part of) the Pauli-Hamiltonian $\mathbb{H }^B $
itself also exists and occurs in one of the sectors with
$\mathbb{M } \geq 1 . $ This was observed in numerical studies.
The ground-state of the $H_2 $ molecule was computed numerically
(by the variational method) in \cite{DSC}, \cite{DSDC}, and that
of the two-electron $He _2 ^{2+ }    $-ion more recently in
\cite{TG}. It was shown that for both these molecules the
ground-state changes, with increasing $B $, from a spin singlet
state with $\mathbb{M } = 0 $ (for $B = 0 $) to a, bound or
unbound, spin triplet state (both electron spins parallel) still
with $\mathbb{M } = 0 $ to a strongly bound spin triplet state
with $\mathbb{M } = 1 $ (using molecular term symbols, the
transition is $^1 \Sigma _u \to \, ^3\Sigma _u \to \, ^3 \Pi _u
$). A similar phenomenon occurs for other small molecular ions
such as $H_3 ^+ $ and $HeH^+ $, cf. \cite{Tu} and its references.
We should note that it is implicit in the definition of $H_{\delta
} $ that all the electron spins are taken anti-parallel to the
magnetic field (since $H_{\delta } $ effectively operates on the
projection of the full Hilbert space with spin onto the lowest
Landau state - cf. \cite{BD3}) so that the ground state of
$H_{\delta } $ will correspond to a triplet state of the
Pauli-Hamiltonian. For further numerical studies of spectra of
atoms and molecules in strong magnetic fields, see for example the
two conference proceedings \cite{AtMol} and \cite{NeuSt}.
\medskip

Let $H_2 (\delta ) $ denote the one-dimensional $H_2$--molecule as
defined by the $\delta $--Hamiltonian, $H_{\delta } . $ We will
study this molecule for arbitrary nuclear charge $Z $, but will
often single out the case of $Z = 1 $, corresponding to $H_2 $, as
well as the case of $Z = 2 $ which provides an aymptotic
description of the molecular $He _2 ^{2+ } $--ion in the large
field limit.
\medskip

The two main results of our paper are:

\begin{theorem} \label{Existence} (Existence of $H_2 (\delta ) $)
\smallskip

\noindent (i) If $Z \geq 1 $, then the electronic Hamiltonian
$h_{\delta } $ possesses a ground state for all $a \geq 0 . $
\medskip

\noindent (ii) The $H_2 (\delta ) $-molecule exists (in Born
Oppenheimer sense) for all $Z \geq 1 $, as long as $Z/L \leq 0.297
. $ For $Z = 2 $, this can be sharpened to $2/L \leq 0.458 $ or
$L^{-1 } \leq 0.229 . $
\end{theorem}

We recall that the molecule exists in Born--Oppenheimer sense if
$\inf _a E(a) < \liminf _{a \to \infty } E(a) $, where $E(a) $ is
the infimum of the spectrum of $H_{\delta } = H_{\delta } (a ) $,
and if moreover there exists an $a = a_{\rm eq } $ such that
$E(a_{\rm eq } ) = \inf _a E(a) $ and $E(a_{\rm eq } ) $ is an
eigenvalue of $H_{\delta } (a_{\rm eq } ) $; note that $\inf _a
E(a) < \liminf _{a \to \infty } E(a) $ and continuity of $E(a) $
already imply that the infimum is attained. We will call any
$a_{\rm eq } $ in which $E(a) $ assumes its global minimum an
equilibrium distance of the molecule.

\begin{theorem} \label{thm:eq-asympt} For fixed $\varepsilon $, let
$a_{\rm eq } (\varepsilon ) $ be an equilibrium distance of the
$H_2 (\delta ) $-molecule. Then there exists a constant $c = c(Z)
$ such that
\begin{equation}
a_{\rm eq } (\varepsilon ) \simeq c \sqrt{\varepsilon }, \ \
\mbox{as} \,\, \varepsilon \to 0 .
\end{equation}
\end{theorem}
The constant is the same for any of the, potentially multiple,
equilibrium distances; it is natural to conjecture that $a_{\rm eq
} (\varepsilon ) $ is in fact unique, as is observed numerically:
see for example figure \ref{fig-4} below.
\medskip

We conjecture that, as a corollary of these theorems, the actual
molecule as modelled by the Pauli Hamiltonian can also be shown
rigorously to exist for $Z \geq 1 $ and sufficiently large $B $,
and that its equilibrium distance behaves as $C (\log B )^{-3/2 }
$ for some constant $C $ (note that by (\ref{def:a;epsilon}), $R
\simeq L^{-3/2 } $ if $a \simeq c \sqrt{\varepsilon} \, $). For
$H_2 ^+ $ this was shown to be the case in \cite{BBDPV1}. We
recall that for $B = 0 $ and $Z = 1 $, the existence of the $H_2
$-molecule was rigorously established in \cite{F-et-al}. A
discussion of the stability scenario for varying $Z $, still with
vanishing magnetic field, can be found in \cite{A-et-al-1},
\cite{A-et-al-2}, where it is in particular shown that $He _2 ^{2+
}    $ is meta-stable when $B = 0 . $ We also note that reduction
of the size of the molecule with increasing magnetic field
strength has been observed numerically: cf. for example \cite{Tu}
and its references.
\bigskip

\section{\bf Existence of the molecule}

The ground-state energy of the one-electron Hamiltonian $h^{(1)}
_{\delta } := -\frac{1 }{2 } \Delta _z + \sum _{\pm } \delta (z
\pm a ) $ on $L^2 (\mathbb{R } ) $, which is defining the
electronic energy of the $H_2 ^+ (\delta ) $-ion, was determined
in \cite{BBDP} as being $-\frac{1 }{2 } \alpha _0 (a)^2 $, where
\begin{equation}
\alpha _0 (a) := 1 + \frac{W\left( 2a e^{-2a } \right) }{2a } ,
\end{equation}
with normalized eigenfunction $\varphi _0 $ given by
\begin{equation}
\varphi_0 (z) = \left \{ \begin{array}{ll} A_1 e^{-
\alpha _0 \, |z | } , \ \ &|z| > a  \\
A_2 \cosh (\alpha _0 z ) , \ \ &|z | < a ,
\end{array}
\right.
\end{equation}
where
\begin{equation}
A_1 = \frac{\sqrt{\alpha _0 } \left( 1 + e^{2 a \alpha _0 }
\right) }{\sqrt{2 } \sqrt{\left( 1 + 2 e^{2 a \alpha _0 } + 2 a
\alpha _0 \right) } } , \ \ A_2 = \frac{\sqrt{2 a \alpha _0 }
}{\sqrt{ \left( 1 + 2 e^{2 a \alpha _0 } + 2 a \alpha _0 \right) }
} .
\end{equation}
In fact, for $a $'s larger than some $Z $-dependent threshold, the
spectrum of $h_{\delta } ^{(1)} $ consists of two eigenvalues
$\alpha _0 (a) $, $\alpha _1 (a) $ less than 0 and a continuous
spectrum equal to $[0, \infty ) $; for $a $'s below this threshold
the excited state is absorbed in the continuous spectrum, cf.
\cite{AM} and \cite{H}. The following properties of $\alpha _0 (a)
$ will be repeatedly used below: $\alpha _0 (a) $ is a decreasing
function of $a $, $\alpha _0 (0) = 2 $ and $\alpha _0 (a) \to 1 $
as $a \to \infty . $ Finally,
\begin{equation}
\alpha _0 (a) \sim 2 - 4a + O(a^2 ) , \ \ a \to 0 .
\end{equation}
\medskip

Recall that the two-electron Hamiltonian $h_{\delta } = h_{\delta
} (a, Z ) $ defined by (\ref{h-delta}) corresponds to the
electronic part of the energy of our $H_2 ({\delta } ) $-molecule.
Let $e(a)=e(a, Z ) $ be the infimum of the spectrum of $h_{\delta
} (a, Z ) $, and
\begin{equation}
E(a) := E(a, Z, \varepsilon ) := e(a , Z ) + \frac{\varepsilon
}{2a } ,
\end{equation}
the ground-state energy of the molecule at fixed internuclear
distance\footnote{as a notational convention, we denote electronic
Hamiltonians and energies by lower case letters $h_{\delta },
e(a), e^{\rm UB} (a) $, etc. and the corresponding molecular
entities (obtained by adding $\varepsilon / 2a $) by upper case
letters $H_{\delta } , E(a), E^{\rm UB } (a) $, etc.}. To prove
that $e(a) $ is an eigenvalue we have to show, according to the
HVZ-theorem, that $e(a) $ is strictly less than the inf of the
spectrum of $h_{\delta } ^{(1) } $, that is:
\begin{equation} \label{HVZ}
e(a) < - \frac{1 }{2 } \alpha _0 (a)^2 .
\end{equation}
We can derive a simple variational upper bound for $e(a) $ by
using $\varphi _0  \otimes \varphi _0 $ as test-function. Explicit
evaluation of
$$
\left \langle \, \delta (z_1 - z_2 ) , \varphi _0 ^2(z_1) \varphi
_0 ^2 (z_2 ) \, \right \rangle = \int _{\mathbb{R } } \varphi _0
(z)^4 dz ,
$$
then leads to the upper bound
\begin{equation} \label{eq:eub}
e^{\rm UB} (a, Z ) := - \alpha _0 (a)^2 + Z^{-1 } f(a)
\end{equation}
where
\begin{equation} \label{eq:f1}
f(a) = \alpha _0 \cdot \frac{8 \cosh ^4 (a \alpha _0 ) + \sinh (4
a \alpha _0 ) + 8 \sinh (2 a \alpha _0 ) + 12 a \alpha _0 }{4
(e^{2 a \alpha _0 } + 2 a \alpha _0 + 1 )^2 } ,
\end{equation}
which can also be expressed as
\begin{equation} \label{eq:f2}
f(a) = \alpha _0 \cdot \frac{e^{4 a \alpha _0 } + 4 e^{2 a \alpha
_0 } + 4 \sinh (2 a \alpha _0 ) + 12 a \alpha _0 + 3 }{4 (e^{2 a
\alpha _0 } + 2 a \alpha _0 + 1 )^2 } .
\end{equation}

It is elementary to show that $f(a) \leq \frac{1 }{2 } \alpha _0
(a) $, the inequality being sharp for $a = 0 . $ The simple upper
bound (\ref{eq:eub}) suffices to show that the ground state of
$h_{\delta } (a) $ exists for any $a \geq 0 $:
\medskip

\noindent {\it Proof of theorem \ref{Existence}(i).} It suffices
to show that $e^{\rm UB } (a, Z ) \leq -\frac{1 }{2 } \alpha _0
(a)^2 $ for $Z \geq 1 $. But this follows from $e^{\rm UB } (a, Z
) \leq - \alpha _0 (a) ^2 + \alpha _0 (a) / 2Z \leq - \alpha _0
(a) ^2 + \alpha _0 (a) / 2 < - \alpha _0 (a) ^2 / 2 $, since
$\alpha _0 (a) > 1 $ for $a \geq 0 . $ \hfill $\Box $
\medskip

To show existence of the molecule, we examine when $\min_a E^{\rm
UB } (a) < \liminf _{a \to \infty } E(a ) $, where
\begin{equation} \label{eq:EUB}
E^{\rm UB } (a) := E^{\rm UB } (a, Z, \varepsilon ) := e^{\rm UB }
(a) + \frac{\varepsilon }{2 a } ,
\end{equation}
the variational upper bound for the molecule's energy. We note the
trivial lower bound:
\begin{equation} \label{eq:Emin}
E(a, Z, \varepsilon ) \geq E(a, \infty , \varepsilon ) = - \alpha
_0 (a) ^2 + \frac{\varepsilon }{2a } =: E^{\rm NI } (a,
\varepsilon ) ,
\end{equation}
where ``NI'' stands for non-interacting electrons.
\medskip

Clearly, $\liminf _{a \to \infty } E(a) \geq $ $\lim _{a \to
\infty } E^{\rm NI } (a) = \lim _{a \to \infty } - \alpha _0 (a)
^2 = -1 $, so to show existence of the molecule it suffices to
show that there exists an $a \geq 0 $ such that  $E^{\rm UB } (a,
Z , \varepsilon ) < -1 $, or, equivalently, $\varepsilon < j(a, Z
) $, where
\begin{equation}
j(a, Z ) := 2 a \left( \alpha _0 (a)^2 - 1 - Z^{-1 } f(a) \right)
.
\end{equation}
To derive a result which is valid for all $Z \geq 1 $, we note
that $Z \geq 1 $ implies that $j(a, Z ) \geq j(a, 1 ) $. The
molecule will therefore exist for any value of the parameter
$\varepsilon = Z/L $ which is less that $\max _a j(a, 1 ) $, for
any $Z\geq 1 $. For the case of $Z = 2 $, we have to have that
$\varepsilon < \max _a j(a, 2 ) $. Part (ii) of theorem
\ref{Existence} is then an immediate consequence of the following
lemma:

\begin{lemma} \label{lemma:max-j} $j(a, 1 ) $ has a global maximum of 0.297
on $a \geq 0 $, which is attained for $a = 0.254 $. The maximum of
$j(a, 2 ) $ is equal to 0.458, and is attained for $a = 0.337 . $
\end{lemma}

\begin{figure}[htp]
\centering
\includegraphics[width=4in,angle=0]{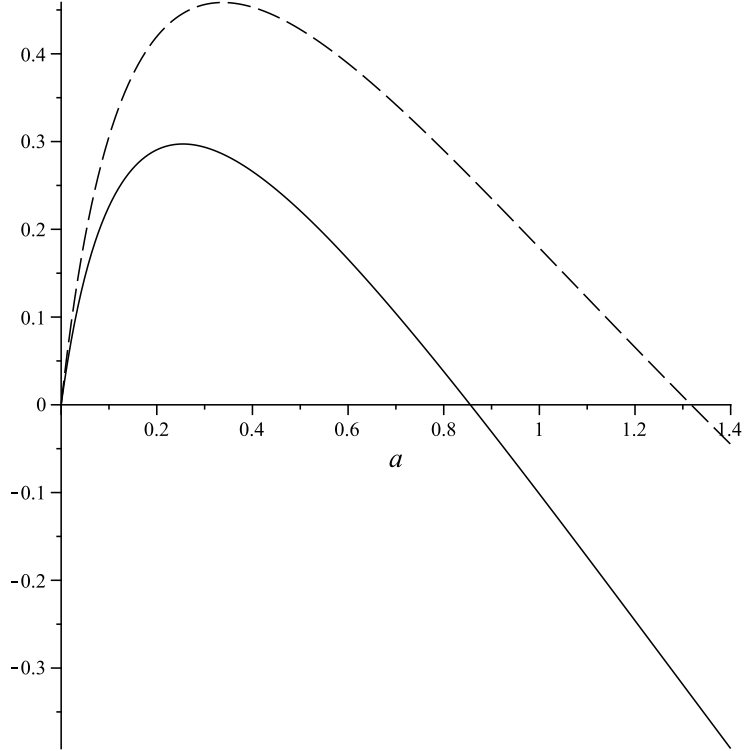}
\caption{Graphs of $j(a, 1) $ (---) and $j(a, 2 ) $ (- - -)}
\label{fig-1}
\end{figure}

This lemma has been checked numerically, using Maple: see figure
\ref{fig-1} for the graphs of $j(a, 1 ) $ and $j(a, 2 ) . $

\begin{remarks} \rm{(i) Our reliance, here and below, on numerical analysis
may be felt to be somewhat unsatisfactory, and it may be possible,
with sufficient effort, to give a rigorous analytic proof of lemma
\ref{lemma:max-j}, since $j(a, Z ) $ has an explicit analytic
expression (and similarly for $g(a, Z ) $ below). This lemma of
course concerns only one particular variational bound (using a
simple test-function) which is unlikely to be optimal, although
numerically it does not perform badly (cf. figure \ref{fig-4}
below). It would also be interesting to find better variational
bounds, if possible again with closed analytic expressions.}
\medskip

\noindent (ii) The Taylor expansion of $j(a, Z ) $ in $a = 0 $
starts of as $j(a, Z ) / 2 = (3 - Z^{-1 } ) a + O(a^2 ) $, so that
$\max _a j(a, Z ) $ will be strictly positive if $Z > 1/3 $.
Graphical analysis shows that $j(a, Z) \leq 0 $ when $Z = 1/3 $
and therefore for all $Z \leq 1/3 $ (note that $\lim _{a \to
\infty } j(a, Z ) = - \infty $). These remarks imply that the
present method would prove existence of the molecule for all $Z >
1/3 $, for sufficiently large $L(B) $, provided we can show that
$h_{\delta } (a, Z ) $ possesses a ground state for such $Z $, at
least for those $a $ for which $j(a, Z ) > 0 $.
\end{remarks}
\medskip

To determine $\min _a E^{\rm UB } (a, Z ) $, we have to solve
$\partial _a e^{\rm UB } = \varepsilon / 2a^2 $, or
\begin{equation} \label{eq:g}
g(a, Z ) = \varepsilon,
\end{equation}
where $g(a, Z ) := 2a^2 \left( - 2 \alpha _0 (a) \alpha _0 ' (a) +
Z^{-1 } f' (a) \right) $, the prime indicating differentiation.
Graphical analysis of $g(a, Z) $ using Maple or Mathematica shows
that (\ref{eq:g}) has two solutions as long as $\varepsilon <\max
_a g(a, Z ) $ - see figure \ref{fig-2} below.
\begin{figure}[htp]
\centering
\includegraphics[width=4in,angle=0]{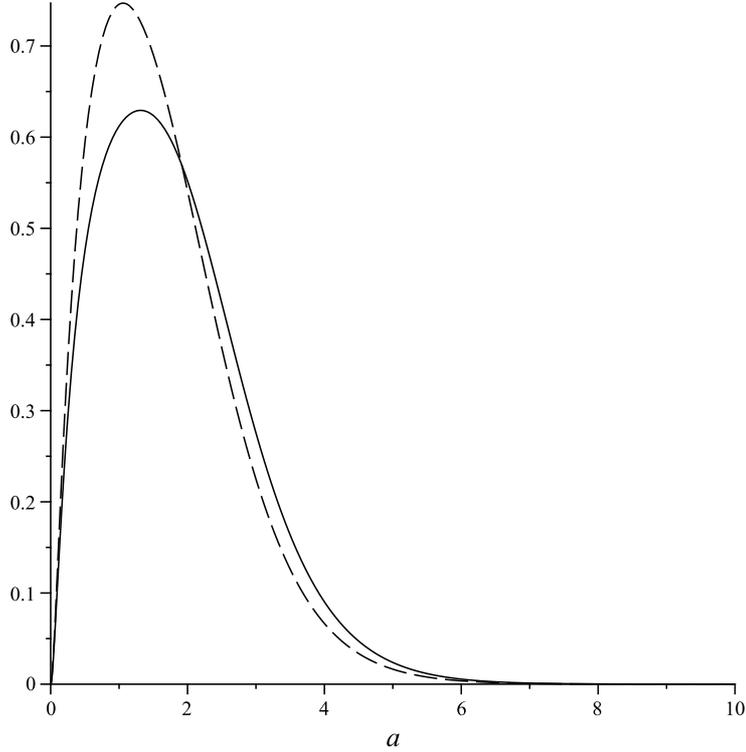}
\caption{Graphs of $g(a, 1) $ (---) and $g(a, 2 ) $ (- - -)}
\label{fig-2}
\end{figure}

\noindent It follows (again only graphically, for the time being)
that $E^{\rm UB } (a, Z, \varepsilon ) $ has two critical points,
of which the smaller one, for sufficiently small $\varepsilon $,
turns out to be a local minimum and the larger a local
maximum\footnote{as follows by further numerical analysis: for
example, in the stationary points of $E^{\rm UB} (a) $, we have
that $(E^{\rm UB } )'' (a) = - 2 \alpha _0 ' (a)^2 - 2 \alpha _0
(a) \alpha _0 '' (a) + Z^{-1 } f''(a) + a^{-3 } g(a, Z ) $, and
the latter function is positive for $a $ in a neighborhood of 0
(e.g. for $a < 2 $ in case $Z = 2 $)}. The local minimum is of
course a global one under the hypotheses of theorem
\ref{Existence}(ii).
\medskip

Let the minimum of $E^{\rm UB } (a, Z, \varepsilon ) $ be attained
in $a = a_{\rm eq } ^{\rm UB } (\varepsilon ) $; $a_{\rm eq }
^{\rm UB } (\varepsilon ) $ would be the equilibrium position of
the molecule if its energy were in fact equal to our upper bound.
It then follows (from the graph of $g(a, Z ) $) that $a_{\rm eq }
^{\rm UB } (\varepsilon ) \to 0 $ as $\varepsilon \to 0 $. This
information turns out to be sufficient to determine its
small-$\varepsilon $ asymptotics, using a simple argument which we
will also use in the next section for the true equilibrium
position of $H_2 (\delta ) $-molecule:

\begin{lemma} \label{lemma:eq-asympt-UB} As $\varepsilon \to 0 $,
\begin{equation} \label{eq-asympt-UB}
a_{\rm eq } ^{\rm UB } (\varepsilon ) \simeq c \sqrt{\varepsilon }
,
\end{equation}
with constant $c $ equal to
\begin{equation}
c = \frac{1 }{\sqrt{2(e^{\rm UB }) \, '(0) } } = \frac{1 } {2 }
\sqrt{\frac{Z }{8Z - 1 } } .
\end{equation}
\end{lemma}

\noindent {\it Proof.} We fix $Z $ and write $e^{\rm UB } (a, Z) =
e^{\rm UB }(a) $. Then $a_{\rm eq } ^{\rm UB } (\varepsilon ) $ is
the smallest of the two solutions, $a = a_- (\varepsilon ) $, of
$(e^{\rm UB })' (a ) = (2a^2 )^{-1 } \varepsilon $, so that
\begin{equation} \nonumber
a_- (\varepsilon ) = \sqrt{\frac{\varepsilon }{ 2 (e^{\rm UB }
)'(a_- (\varepsilon )) } } .
\end{equation}
Now, as we observed above, $a_- (\varepsilon ) \to 0 $ as
$\varepsilon \to 0 $, and therefore $(e^{\rm UB } )' (a_-
(\varepsilon )) \to (e^{\rm UB } )'(0) $, from which the first
asymptotic equality of (\ref{eq-asympt-UB}) follows. For the
second equality, we use that $e^{\rm UB } (a) = -(4 - Z^{-1 } ) +
(16 - 2 Z^{-1 } )a + O(a^2 ) . $ \hfill $\Box $
\medskip

Note that the argument we gave here is very general, and we will
seek to apply it in the next section to the true equilibrium
position, $a_{\rm eq } (\varepsilon )$ of $H_2 (\delta ) . $
\medskip

The equilibrium position $a_{\rm eq } ^{\rm UB } (\varepsilon ) $
we just studied has of course no real physical significance. We
will use it, though, in the next section, to obtain a weak
a-priori estimate for $a_{\rm eq } (\varepsilon ) . $ To that
effect, we note that if we combine (\ref{eq-asympt-UB}) with the
Taylor expansion of $E^{\rm UB } (a, \varepsilon ) $ we find:

\begin{corollary} \label{Corr:minE^UB}There exists a constant $C > 0 $ such
that
\begin{equation} \label{minE^UB}
E^{\rm UB } _{\rm min } (\varepsilon ) := \min _a E^{\rm UB } (a,
\varepsilon ) \simeq -4 + Z^{-1 } + C \sqrt{\varepsilon } , \ \
\varepsilon \to 0 .
\end{equation}
In particular, $E^{\rm UB } _{\rm min } (\varepsilon ) < -1 $ for
$Z \geq 1 $ and $\varepsilon $ sufficiently small .
\end{corollary}

One can perform a similar analysis for the lower bound,
\begin{equation} \nonumber
E^{\rm NI } (a, \varepsilon ) = - \alpha _0 (a)^2 +
\frac{\varepsilon }{2a } ,
\end{equation}
and show that the curve $E^{\rm NI } (a, \varepsilon ) $ has, for
fixed and small enough $\varepsilon $, a local minimum and a local
maximum, the local minimum again being absolute for sufficiently
small $\varepsilon $; the local maximum lies above $\lim _{a \to
\infty } E^{\rm NI }(a, \varepsilon ) = - 1 $. If $a_{\rm eq }
^{\rm NI } (\varepsilon ) $ is the location of the minimum, then
we have again that $a_{\rm eq } ^{\rm NI } (\varepsilon ) \simeq c
\sqrt{\varepsilon } $ with a constant $c $ which is now equal to
$c = (2 (e^{\rm NI } )'(0) )^{-1/2 } $, and
\begin{equation} \label{minE^NI}
\min _a E^{\rm NI } (a , \varepsilon ) \sim -4 + C
\sqrt{\varepsilon } ,
\end{equation}
for a suitable constant $C . $ We finally observe, from the
geometry of the graph of $E^{\rm NI } (\cdot , \varepsilon ) $,
that $E^{NI } (a) $ will be strictly increasing on any interval
$[a_{\rm eq } ^{\rm NI } (\varepsilon ) , A ] $ such that $E^{\rm
NI } (A , \varepsilon ) < -1 = \lim _{a \to \infty } E^{\rm NI }
(a) . $

\section{\bf Asymptotics of $a_{\rm eq } (\varepsilon ) $}

Recall that $e(a) := e(a, Z ) $ is the ground-state energy of
$h_{\delta } $, and that $E(a, \varepsilon) = e(a) + \varepsilon /
2a $. As we have seen, $E(a, \varepsilon ) $ possesses an absolute
minimum if $\varepsilon = Z/L $ is sufficiently small. It is not a
priori known whether this minimum is unique, though we would
expect it to be. Below, we let $a_{\rm eq }(\varepsilon ) $ be any
value of $a $ at which $E(a, \varepsilon ) $ attains its absolute
minimum in $a $ on $[0, \infty ) . $

\medskip

We will use the following lemma on differentiability of $e(a) $,
whose proof we postpone till the the end of this section, in order
not to interrupt the flow of the argument.

\begin{lemma} \label{diff-e(a)} The ground-state energy $e(a) $ is continuously differentiable on
$[0, \infty ) . $ In particular, its right-derivative $e'(0^+ ) $
in 0 exists. Moreover, $e'(0^+ ) > 0 $ for $Z > 1/4 . $
\end{lemma}

The next lemma takes up the idea of lemma \ref{lemma:eq-asympt-UB}
but now for $e(a) $ directly.

\begin{lemma} \label{lemma:eq-asympt} Let $a = a(\varepsilon ) $ be a solution of
$e'(a ) = \varepsilon / 2 a^2 $ for which there exists a constant
$A
> 0 $ such that
\begin{enumerate}
\item $a(\varepsilon ) \in [0, A ] $ for $\varepsilon $
sufficiently small;

\item $\min _{a \in [0, A ] } e'(a) > 0 . $
\end{enumerate}
Then as $\varepsilon \to 0 $,
\begin{equation}
a_{\rm eq } (\varepsilon ) \simeq \sqrt{\frac{\varepsilon }{2
e'(0^+ ) } } .
\end{equation}
\end{lemma}

\noindent {\it Proof.} Since $2 a(\varepsilon ) ^2 = \varepsilon /
e'(a(\varepsilon ) ) \leq \varepsilon / \min _{[0, A ] } e'(a) \to
0 $ as $\varepsilon \to 0 $, it follows that $e'(a(\varepsilon ) )
\to e'(0+) $, and hence that $2 a(\varepsilon )^2 / \varepsilon
\to e'(0+) ^{-1 } . $ \hfill $\Box $
\medskip

We next use $E^{\rm NI } (a, \varepsilon ) $ and $E^{\rm UB }
_{\rm min } (\varepsilon ) $ (cf. lemma \ref{Corr:minE^UB}) to
effectively bound $a_{\rm eq } (\varepsilon ) $ (compare
\cite{BBDPV1}, proof of theorem 2). We assume $\varepsilon $
sufficiently small for $E^{\rm UB } (\varepsilon ) < -1 $ to hold:
cf. corollary \ref{Corr:minE^UB}.

\begin{lemma} \label{lemma:A_+} Suppose that $\varepsilon $ is sufficiently small so that
$E_{\rm min } ^{\rm UB } (\varepsilon ) < -1 . $ Let $A = A_+
(\varepsilon ) $ be the largest of the two roots of $ E^{\rm NI }
(A, \varepsilon ) = E_{\rm min } ^{\rm UB } (\varepsilon ) . $
Then $a_{\rm eq } (\varepsilon ) \leq A_+ (\varepsilon ) . $
\end{lemma}

\noindent {\it Proof.} We first claim that $E^{\rm NI } (a_{\rm eq
} (\varepsilon ), \varepsilon ) < -1 $. For otherwise, $E ( a_{\rm
eq } (\varepsilon ), \varepsilon ) \geq E^{\rm NI } (a_{\rm eq }
(\varepsilon ), \varepsilon ) \geq -1 $ which is in contradiction
with $E^{\rm UB } (\varepsilon) < -1 $ (since $E_{\rm min } ^{\rm
UM } (\varepsilon) $ dominates the ground-state energy of
$H_{\delta } $). Suppose now that $a_{\rm eq } (\varepsilon ) $ is
strictly larger than $A_+(\varepsilon ) $ which, as the larger
root, is bigger than $a_{\rm eq } ^{\rm NI } (\varepsilon ) . $ It
then follows from the properties of $a \to E^{\rm NI } (a,
\varepsilon) $ that $E^{\rm NI } (a, \varepsilon ) $ will be
strictly increasing on the interval $[a_{\rm eq } ^{\rm NI }
(\varepsilon ) , a_{\rm eq } (\varepsilon) ] $, and hence $
E(a_{\rm eq } (\varepsilon ) , \varepsilon ) \geq E^{\rm NI }
(a_{\rm eq } (\varepsilon ) , \varepsilon ) > E^{\rm NI } (A_+
(\varepsilon ) , \varepsilon ) = E^{\rm UB } _{\rm min }
(\varepsilon ) \geq 
E( a_{\rm eq } (\varepsilon ), \varepsilon ) $, which is a
contradiction. \hfill $\Box $
\medskip

One might hope that the previous lemma, combined with lemma
\ref{lemma:eq-asympt-UB}, would already imply that $a_{\rm eq }
(\varepsilon ) \to 0 $, in which case lemma \ref{lemma:eq-asympt}
together with $e'(0+) > 0 $ would already imply theorem
\ref{thm:eq-asympt}. However, this is unfortunately not the case:
if $A_+ (\varepsilon ) \to 0 $, then since $A_+ (\varepsilon )
\geq a_{\rm eq } ^{\rm UB } (\varepsilon ) \sim \sqrt{\varepsilon}
$,
$$
E^{\rm NI } (A_+ (\varepsilon ) , \varepsilon ) \leq \alpha _0
(A_+ (\varepsilon ) )^2 + c \sqrt{\varepsilon } \to \alpha _0
(0)^2 = -4 ,
$$
but on the other hand $E^{\rm NI } (A_+ (\varepsilon ) ,
\varepsilon ) = E^{\rm UB }_{\rm min } (\varepsilon ) \to - 3 +
Z^{-1 } $, by (\ref{minE^UB}), which is a contradiction. However,
it is easy to show that $a_{\rm eq } (\varepsilon ) = O(1) $ as
$\varepsilon \to 0 $, since if $A_+ (\varepsilon _{\nu } ) \to
\infty $ on some sequence $\varepsilon _{\nu } \to 0 $, then
$$
E^{\rm NI } (A_+ (\varepsilon _{\nu } ), \varepsilon _{\nu } ) = -
\alpha _0 (A_+ (\varepsilon _{\nu } ) )^2 + \frac{\varepsilon
_{\nu } }{ A_+ (\varepsilon _{\nu } ) } \to -1 ,
$$
which contradicts
$$
E^{\rm NI } (A_+ (\varepsilon _{\nu } ), \varepsilon _{\nu } ) =
E^{\rm UB }_{\rm min } (\varepsilon _{\nu } ) \simeq - 4 + Z^{-1 }
+ C \sqrt{\varepsilon _{\nu } } \to -4 + Z^{-1 } ,
$$
as long as $Z > 1/3 . $ {\it If we now would know that $e'(a) > 0
$ for all $a \geq 0 $}, this O(1)-estimate for $a_{\rm eq }
(\varepsilon ) $ in combination with lemma \ref{lemma:eq-asympt}
would prove theorem \ref{thm:eq-asympt}. We were however not able
to prove monotonicity of $e(a) $. What we can do is give an
effective upper bound for $a_{\rm eq } (\varepsilon ) $, which
means we only have to check monotonicity numerically on some known
finite interval.

\begin{lemma} \label{interval} Let $Z \geq 1 $ and let $\varepsilon $ be such that $E^{\rm UB } _{\rm
min } (\varepsilon ) \leq -2 $: (this is true for all sufficiently
small $\varepsilon$ by Corollary \ref{Corr:minE^UB}). Then $A_+
(\varepsilon ) \leq  \alpha _0 ^{-1 } (\sqrt{2} ) = 0.3116 . $
Consequently, $a_{\rm eq } (\varepsilon ) \leq 0.3116 . $
\end{lemma}

\noindent {\it Proof.} Elementary, if we draw the graphs of $E^{NI
} (a, \varepsilon ) = - \alpha _0 (a)^2 + \varepsilon / 2a $, of
$-\alpha _0 (a)^2 $, which is an increasing function from -4 in 0
to -1 at infinity, and of the constant functions $E^{\rm UB }
_{\rm min } (\varepsilon ) $ and $-2 . $ \hfill $\Box $
\medskip

\begin{figure}[htp]
\centering
\includegraphics[width=5in]{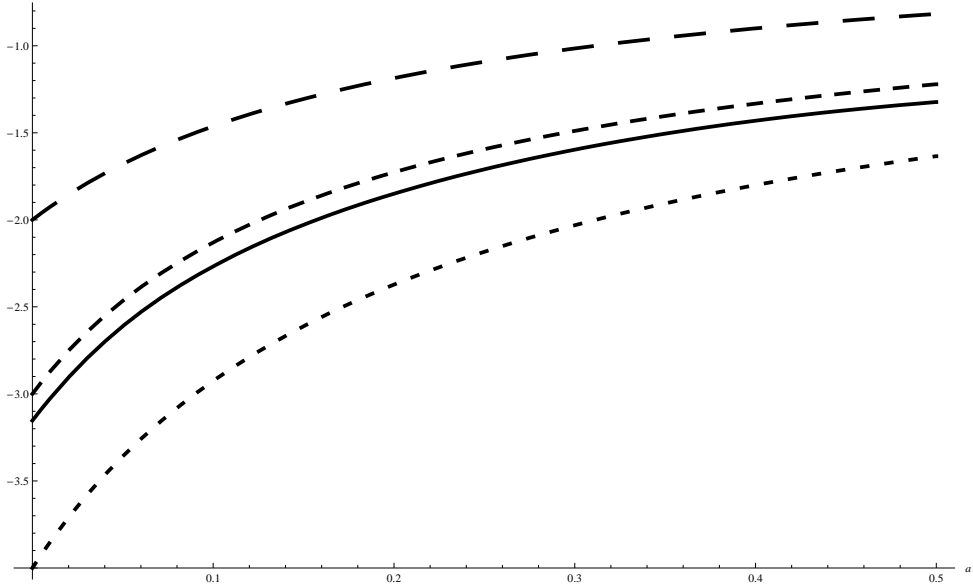}
\caption{Electronic energy curves; from top to bottom: $-\frac{1
}{2} \alpha _0 (a)^2 $ ({\bf -- -- -- }); $e^{\rm UB} (a) \, $(- -
- ); $e(a) \, $(\bf{------}); $e^{\rm NI }(a) \, $({\tiny - - -
})}\label{fig-3}
\end{figure}

We can therefore take $\alpha _0 ^{-1 } (\sqrt{2} ) $ in lemma
\ref{lemma:eq-asympt} (smaller $A $'s are also possible, e.g. for
$Z = 2 $, depending on how large we are willing to let $Z$ be or
how small $\varepsilon $). To finish the proof of theorem
\ref{thm:eq-asympt} we verify numerically that $e(a) $ is strictly
increasing on $[0, \alpha _0 ^{-1 } (\sqrt{2} ) ] $. This can be
done using Pierre Duclos' skeleton method to compute $e(a) $, and
has been carried out by one of us in \cite{B}.  The plot of $e(a)
$ reveals its monotonicity over the desired interval, thereby
completing the proof of theorem \ref{thm:eq-asympt}: see figure
\ref{fig-3}, which for comparison also shows the one-electron
energy $-\frac{1 }{2 } \alpha _0 (a)^2 $, as well as $e^{\rm UB }
(a) $ and $e^{\rm NI } (a) . $ We note in particular, here and
also in figure 4 below, that our variational upper bound, though
relatively naive, gives a quite reasonable approximation to the
actual ground-state energy.
\medskip

\begin{figure}
\begin{center}
  \includegraphics[scale=0.6]{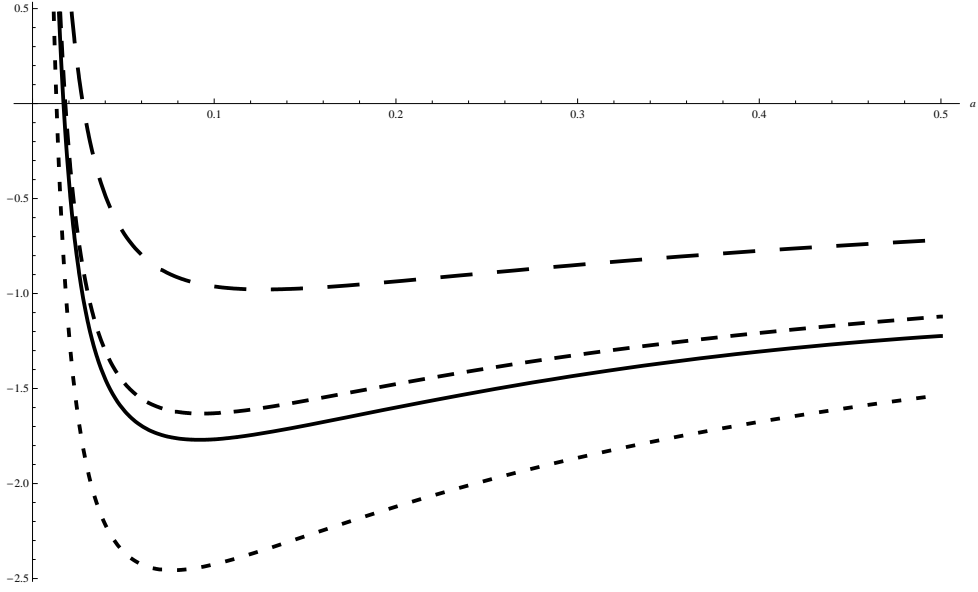}\\
  \caption{Molecular energy curves; from top to bottom: $-\frac{1
}{2} \alpha _0 (a)^2 + \varepsilon / 2a $ ({\bf -- -- -- });
$E^{\rm UB} (a) \, $(- - - ); $E(a) \, $(\bf{------}); $E^{\rm NI
}(a) \, \newline $({\tiny - - - })}\label{fig-4}
  \end{center}
\end{figure}

The skeleton method allows us to numerically compute the
equilibrium distance and energy of the $H_2 (\delta ) $-molecule
for a given $Z $ and $L $. We discuss by way of example the case
of a Hydrogen molecule ($Z = 1 $) in a magnetic field with
field-strength corresponding to $L = 10 $, or $B = L^2 e^L \simeq
2.2 \, 10^6 $. We recall (cf. for example [BBDP]) that the
magnetic field is measured in units of $B_0 = \frac{m_e^3 e^3
c}{\hbar^3}=2.35\times 10^{9} G = 2.35\times 10^{5} T$, where $G$
stands for Gauss and $T$ for Tesla. Moreover, the energy is
measure in units of $\hbar\omega_0=\hbar\frac{e B_0}{m_e c}=27.2
eV=1 \, \mathrm{Hartree}$, where $\omega_0$ is the cyclotron
frequency, and distance in units of the Bohr radius
$a_0=\frac{\hbar^2}{m_e e^2}=0.53 {\AA}$. Our $L = 10 $ therefore
corresponds to a magnetic field of $5.17 \, 10^{11 } \, T $, which
is of course not realizable on earth, but may be realistic for a
neutron star. Figure \ref{fig-4} displays the graph $E(a) $ (solid
line) as well as, for the sake of comparison, that of $E^{\rm UB }
(a) $ (medium dash line), $E^{\rm NI } (a) $ (small dash line) and
$-\frac{1 }{2 } \alpha_0(a )^2 + \epsilon/2a $ (large dash line).
We see that $a_{eq}\approx 0.1$ and $E_{eq}\approx-1.75$, which
corresponds to an equilibrium energy of $\hbar\omega_0 \times
-1.75 = -47.6 eV = -1.75 \, \mathrm{Hartree}$ and an equilibrium
distance of $R_{eq}=a_0\times\frac{2 a}{L Z}\approx 10^{-2} {\AA}
. $ The equilibrium distance is much smaller than $a_0$ but still
significantly bigger than the distance of $\approx 10^{-5} {\AA}$
over which the nuclear interaction between two protons makes
itself felt. Nevertheless, following \cite{AH}, it may be small
enough to significantly enhance the probability for protons to
pass through the electronic barrier and be trapped in the nuclear
well. It would be interesting to compute the tunnelling
cross-section in the Gamov model for the $H_2 (\delta ) $-model,
as simplified model for the actual $H_2 $ molecule.
\medskip

We finish with the promised proof of lemma \ref{diff-e(a)}.
\medskip

\noindent {\it Proof of lemma \ref{diff-e(a)}}. We will carry out
the proof for the $N $-electron Hamiltonian
\begin{equation}
h_{\delta } (a) := \sum _i \left( - \frac{1 }{2 } \Delta _i - \sum
_{\pm } \delta (z_i \pm a ) \right) + \frac{1 }{Z } \sum _{i < j }
\delta (z_i - z_j ) ,
\end{equation}
where we will suppose, to fix ideas, that $a \geq 0 $ (though this
is not strictly speaking necessary). Differentiability of $e(a) $
follows from norm-differentiability of the resolvent, which can be
established using the symmetrized resolvent equation - we skip the
details.
\medskip

We next study $e'(a) $. Let $\psi _a $ be the normalized
ground-state eigenfunction of $h_{\delta } (a) $ on $L^2
(\mathbb{R }^N ) $, which is unique. A formal application of the
Feynman-Hellman theorem would lead to
\begin{equation} \nonumber
e'(a) = - \sum _i \sum _{\pm } \, (\delta ' (z_i \pm a ) \psi _a ,
\psi _a ) = - \sum _i \sum _{\pm } \, \langle \delta ' (z_i \pm a
) , | \psi _a |^2 \rangle .
\end{equation}
However, we have to be careful here: since $\partial _{z_i } \psi
_a $ a priori has a jump in $z_i = \pm a $, $|\psi _a |^2 $ is not
an admissible test-function for $\delta ' (z_i \pm a ) . $

We will first establish Feynman-Hellman on the level of quadratic
forms. If we write $\psi _a (z) = \psi (z, a ) $ (to more clearly
bring out the parameter dependence in the notations), the
eigen-equation $h_{\delta } (a) \psi _a = e(a) \psi _a $
translates into
$$
\frac{1 }{2 } (\nabla \psi (\cdot , a ) , \nabla \varphi ) - \sum
_i \sum _{\pm } \int _{z_i = \pm a } \, \psi (\cdot , a ) \,
\overline{\varphi } + Z^{-1 } \sum _{i < j } \int _{z_i = z_j } \,
\psi (\cdot , a ) \, \overline{\varphi } = e(a) (\psi (\cdot , a )
, \varphi ) ,
$$
for all $\varphi $ in $H^1 (\mathbb{R }^N ) $, the form-domain of
$h_{\delta } (a) $ We now carefully differentiate with respect to
$a $, taking into account that $\psi (\cdot , a ) $ is only left-
and right- differentiable with respect to $z_i $ in $z_i = \pm a
$. Furthermore, the derivative of the arbitrary $H^1 $-function
$\varphi $ only exists in $L^2 $-sense. However, anticipating that
we will take $\varphi = \psi _a = \psi (\cdot , a ) $ below, we
assume that the relevant left- and right- derivatives of $\varphi
$ also exist. Using the norm-differentiability of $\psi (\cdot , a
) $ with respect to $a $ (as an $L^2 $-valued function of $a $)
and taking right-derivatives with respect to $a $, we arrive at:
\begin{eqnarray*}
\frac{1 }{2 } ( \nabla \partial _a \psi (\cdot , a ) , \nabla
\varphi ) &- & \sum _i \sum _{\pm } \, \int _{\mathbb{R }^{N - 1 }
} \,
\partial _a \psi (z_1 , \ldots \pm a , \ldots , z_N , a ) \,
\overline{\varphi } (z_1 , \ldots , \pm a , \ldots z_N ) \\
&- & \sum _i \int _{\mathbb{R }^{N - 1 } } \,
\partial _{z_i } \psi (z_1 , \ldots a^+ , \ldots , z_N , a ) \,
\overline{\varphi } (z_1 , \ldots , a , \ldots , z_N ) \\
&- & \sum _i \int _{\mathbb{R }^{N - 1 } } \, \psi (z_1 , \ldots a
, \ldots , z_N , a ) \, \partial _{z_i } \overline{\varphi }
(z_1 , \ldots , a^+ , \ldots , z_N ) \\
&+ & \sum _i \int _{\mathbb{R }^{N - 1 } } \,
\partial _{z_i } \psi (z_1 , \ldots -a^- , \ldots , z_N , a ) \,
\overline{\varphi } (z_1 , \ldots , a , \ldots , z_N ) \\
&+ & \sum _i \int _{\mathbb{R }^{N - 1 } } \, \psi (z_1 , \ldots a
, \ldots , z_N , a ) \, \partial _{z_i } \overline{\varphi } (z_1
, \ldots
, -a^- , \ldots , z_N ) \\
&+ &Z^{-1 } \sum _{i < j } \int _{z_i = z_j } \,
\partial _a \psi (\cdot
, a ) \, \overline{\varphi } \\
&=& e(a) (\partial _a \psi (\cdot , a ) , \varphi ) + e'(a) (\psi
(\cdot , a ) , \varphi ) .
\end{eqnarray*}
(Note that the right-derivative (with respect to $a $) of a term
such as $\psi (z_1 , \ldots , -a , \ldots z_N ) $ is $ - \partial
_{z_i } \psi (z_1 , \ldots , -a^- , \ldots , z_N ) . $) As
announced, we now take $\varphi = \psi (\cdot , a ) $. Writing
again $\psi _a $ for $\psi (\cdot , a ) $ and using
real-valuedness of $\psi _a $ (being the ground state function),
we then arrive at the identity
\begin{eqnarray*}
(h_{\delta } (a) \partial _a \psi _a , \psi _a ) &- & \sum _i \int
_{\mathbb{R }^{N - 1 } } \, 2 \psi _a (z_1 , \ldots , a , \ldots ,
z_N ) \, \partial _{z_i } \psi _a (z_1 , \ldots a^+ , \ldots , z_N
) \\
&+ & \sum _i \int _{\mathbb{R }^{N - 1 } } \, 2 \psi _a (z_1 ,
\ldots , -a , \ldots , z_N ) \, \partial _{z_i } \psi _a (z_1 ,
\ldots -a ^- , \ldots , z_N ) \\
&=& e(a) (\partial _a \psi _a , \psi _a ) + e'(a) (\psi _a , \psi
_a ) .
\end{eqnarray*}
Since $( h_{\delta } (a) \partial _a \psi _a , \psi _a ) =
(\partial _a \psi _a , h_{\delta } (a) \psi _a ) = e(a) (\partial
_a \psi _a , \psi _a ) $, we have proved:

\begin{lemma} (Feynman-Hellman for $h_{\delta } (a) $). Let $\psi _a $ be the
normalized ground-state of $h(a) $. Then
\begin{equation} \label{eq:e'}
e'(a) = \sum _i \left( \, - 2 \int _{z_i = a^+ } \, \psi _a
\partial _{z_i } \psi _a + 2 \int _{z_i = -a ^- } \psi _a \partial
_{z_i } \psi _a \right) .
\end{equation}
\end{lemma}

\noindent {\it End of the proof of lemma \ref{diff-e(a)}. } If we
now let $a \to 0 $ in (\ref{eq:e'}), and use the boundary
conditions at $a = 0 $ for membership of the domain of $h_{\delta
} (0) $, we obtain that
\begin{eqnarray*}
e'(0 ^+) &=& 2 \sum _i \int _{z_i = 0 } \left( \partial _{z_i }
\psi _0 |_{z_i = 0^- } \ - \ \partial _{z_i } |_{z_i = 0^+ }
\right) \psi _0
\\
&=& 4 \sum _i \int _{z_i = 0 } \psi _0 ^2 \geq 0 .
\end{eqnarray*}
Since, finally,
\begin{eqnarray*}
2 \sum _i \int _{z_i = 0 } \psi _0 ^2 &=& \frac{1 }{2 } || \nabla
\psi _0 ||^2 + \sum _{i < j } \left( \delta (z_i - z_j ) \psi _0 ,
\psi _0 \right) - ( \, h_{\delta } (0) \, \psi _0 , \psi _0 \, ) \\
&\geq & - ( \, h_{\delta }(0) \, \psi _0 , \psi _0 \, ) = - e(0) ,
\end{eqnarray*}
we find, specializing to $N = 2 $ again, that $e'(0^+ ) \geq - 2
e(0) \geq - 2 e^{\rm UB } (0) = 2 \alpha _0 (0)^2 - 2 Z^{-1 } f(0)
= 2 \alpha _0 (0) ^2 - Z^{-1 } \alpha _0 (0) > 0 $ as long as $Z >
(2 \alpha _0 (0) )^{-1 } = 1/4 . $ \hfill $\Box $

\section{\bf Conclusions}

We have approximated the $H_2 $-molecule in a constant magnetic
field, as described by the non-relativistic Pauli Hamiltonian with
fixed nuclei, by a two-electron model-Hamiltonian of a
one-dimensional molecule with electron-electron and
electron-nuclei interaction given by $\delta $-potentials, and
interaction between the two nuclei given by the usual Coulomb
potential. It can be shown, using the methods of \cite{BD3}, that
this approximation is exact in the large field limit. We have
shown, for this approximation, that the ground state of the
molecule exists, for any nuclear charge $Z \geq 1 $ and for
sufficiently large magnetic fields, and that the (re-scaled)
inter-nuclear equilibrium distance of the molecule tends to 0 with
increasing field-strength $B $, at a rate of $(\log B )^{-1/2 } .
$ This generalises earlier results for $H_2 ^+ $ in \cite{BBDP},
\cite{BBDPV1}, \cite{BBDPV2}. Of the numerous questions which
remain, the foremost one is to extend these results to the full
Pauli-Hamiltonian. This was possible for the single-electron $H_2
^+ $-ion, but the argument for the equilibrium distance given in
\cite{BBDPV1} used the exact solution of the $\delta $-model. In
the case of the two-electron $H_2 $ such exact solutions are not
available anymore, despite the apparent simplicity of the $\delta
$-potentials, and we have had in part to rely on numerical
computations to prove our results, notably to establish
monotonicity of the electronic ground-state energy as function of
the inter-nuclear distance. Since this would already imply that
the inter-nuclear distance tends to 0 at the proper rate (cf. the
remarks just before lemma \ref{interval}), it would be very
interesting to find an analytical proof of this monotonicity, or
at least on a sufficiently large interval (such as the one
specified in lemma \ref{interval}). Another interesting question
which remains open is that of uniqueness of the equilibrium
position, for the $\delta $-approximation as well as for the full
Pauli-molecule.

We also relied on numerical computations in the study of our
variational upper bound and lower bounds. Here it may be possible
to give analytic proofs, since these upper and lower bounds have
analytically closed expressions involving a particular and
well-understood special function, the Lambert $W $-function. It
would furthermore be interesting to find sharper variational upper
bounds, e.g. by using test functions which incorporate
electron-electron correlations.

Other perspectives are to go beyond the Born--Oppenheimer
approximation and analyse the effect of nuclear vibrations on the
stability of the nuclei, and to compute the probability of
tunnelling through the Coulomb barrier. Finally, it would be
interesting to see to what extend the methods and results of this
paper can be extended to diatomic molecules with an arbitrary
number of electrons. Some (unpublished) work in this direction was
already started by Pierre Duclos.

\end{document}